\documentclass{elsart}

\usepackage{epsfig,amsfonts,amssymb}

\newtheorem{THEOREM}{Theorem}

                        {\end{THEOREM}}
\newcommand{\ax}{\begin{axiom}}
\newcommand{\eax}{\end{axiom}}

\def\L{\mathcal{L}}
\def\a{\alpha}
\def\b{\beta}

\begin{document}
\begin{frontmatter}
\title
{Is the Lorentz signature of the metric of spacetime electromagnetic
  in origin?}

\author[label1]{Yakov Itin}
\ead{itin@math.huji.ac.il}
\author[label2]{and Friedrich W.\  Hehl}

\address[label1]{Institute of Mathematics, Hebrew University of
  Jerusalem \& Jerusalem College of Engineering, Jerusalem 91904,
  Israel, email: {\tt itin@math.huji.ac.il}}
\address[label2] {Institute for
  Theoretical Physics, University of Cologne, 50923 K\"oln, Germany,
  and Department of Physics and Astronomy, University of
  Missouri-Columbia, Columbia, MO 65211, USA, email: {\tt
    hehl@thp.uni-koeln.de}}


{05 January 2004, {\it file Lenz11.tex}} \\

\begin{abstract}
  We formulate a premetric version of classical electrodynamics in
  terms of the excitation $H=({\mathcal H},{\mathcal D})$ and the
  field strength $F=(E,B)$. A local, linear, and symmetric spacetime
  relation between $H$ and $F$ is assumed. It yields, if
  electric/magnetic reciprocity is postulated, a {\em Lorentzian\/}
  metric of spacetime thereby excluding Euclidean signature (which is,
  nevertheless, discussed in some detail). Moreover, we determine the
  Dufay law (repulsion of like charges and attraction of opposite
  ones), the Lenz rule (the relative sign in Faraday's law), and the
  sign of the electromagnetic energy. In this way, we get a systematic
  understanding of the sign rules and the sign conventions in
  electrodynamics. The question in the title of the paper is answered
  affirmatively.
\end{abstract}
\begin{keyword}
{\it Metric of spacetime, classical electrodynamics,
  signature of metric, Lenz's rule, positivity of energy}

\PACS: 03.50.De, 04.20.Cv, 11.30.Er
\end{keyword}
\end{frontmatter}
\section*{Running title}

{Lorentz signature and electromagnetism}
\newpage

\section*{Contents}
\begin{enumerate}
\item Introduction

\item Premetric electrodynemics and its (1+3)-decomposition

\item Energy-momentum current

\item Spacetime relation

\item Positivity of the electromagnetic energy

\item Field equations --- derivation of factors

\item Lorentz force --- relation between different signatures

\item Main results and discussions

\noindent Acknowledgments

\noindent References
\end{enumerate}

\section{Introduction}

The spacetime structure presently used in physics is provided by
special and by general relativity theory. Locally --- and in special
relativity also globally --- one can introduce suitable coordinates
$x^0=ct,\,x^1=x,\,x^2=y, \,x^3=z$, here $c$ is the vacuum velocity of
light, such that the metric reads
\begin{equation}\label{metric}
ds^2=+c^2dt^2-dx^2-dy^2-dz^2\,.
\end{equation}
The qualitative difference between time, with coordinate $t$, and
space, with coordinates $x,y,z$, is reflected in the Minkowskian (also
known as Lorentzian) $(+--\,-)$ signature of the metric of spacetime.

The form of the metric (\ref{metric}) raises immediately three
questions that are related to one another: (i) Why is space
3-dimensional?  (ii) Why is time 1-dimensional? (iii) Why does this
specific choice of the signs appear in the signature?

Arguments have been provided in favor of the 3-dimensionality of
space, such as the stability of orbits in a Newtonian potential, see
Ehrenfest \cite{Ehrenfest}.  Also the sign difference $+\,-$ in the
signature has been addressed in different contexts. Greensite
\cite{Greensite}, for example, has put forward the idea to start with
the complex signature $(e^{i\theta},1,...,1)$ and to treat $\theta$ as
a quantum field that, under certain circumstances, can take the values
$\theta=0$ or $\pm\pi$. Thereby Euclidean or Lorentzian signatures
could emerge, respectively. These ideas have been extended, see, e.g.,
Carlini and Greensite \cite{Carlini} and Odintsov et al.\ 
\cite{Odintsov}, by studying a quantum evolution equation and its
consistency conditions. Tegmark \cite{Tegmark}, starting {}from
superstring theories and applying arguments related to the anthropic
principle, finds only the $1 + 3$ dimensional spacetime as habitable.
Manko${\rm {\check{c}}}$ Bor${\rm{\check{s}}}$tnik and Nielsen
\cite{Mankoc2000,Mankoc2002} have studied equations of motion for
particles with spin in higher dimensional spaces. Requiring linearity
in momentum, hermiticity, irreducibility under the Lorentz group,
together with other technical assumptions, they could single out $1 +
1$ and $1 + 3$ dimensional spacetimes. This type of argument has been
deepened by van Dam and Ng \cite{vanDam}. They consider 4 dimensions
and base their work on Wigner's unitary irreducible representations of
the (proper orthochronous) Poincar\'e group. Inquiring into the
covering groups of the subgroups SO(4), SO(1,3), and SO(2,2) of the
Poincar\'e type group, they find that a $0 + 4$ world has no
interesting dynamics whereas a $2 + 2$ world can only have spin 0
particles; in contrast, a $1 + 3$ world has a rich dynamics.

Our approach is different. With exception of the Ehrenfest argument
quoted, the other considerations are all of quantum (field)
theoretical nature. Is this really plausible? The light cone $ds^2=0$
is defined physically by propagating classical electromagnetic waves
(``light'') in their geometrical optics limit. Accordingly,
understanding the signature of the metric would appear to be
equivalent to the understanding of the properties of propagating light
in its geometrical optics limit (see \cite{ROH2002}). In other words,
if we somehow could deduce the light cone in the framework of {\it
  classical electrodynamics,\/} then the signature would come jointly
with it, and in this way a relation could be established between the
signature of the metric and the laws of electrodynamics.  We will
show, for the first time, that there is a relation between {\em four}
signs in electrodynamics and the signature: the $-$ sign in the
Amp\`ere-Maxwell law $d{\mathcal H}\ominus \dot {\mathcal D}=j$, the $+$ sign
in the Faraday law $dE\oplus \dot B=0$ (the Lenz
rule{\footnote{Discussions of the physics of Lenz's rule can be found,
    e.g., in Jackson \cite{Jackson} --- he calls it Lenz's law --- or
    in Sommerfeld \cite{Sommerfeld}.}}), the $+$ sign of the
electromagnetic energy density, and the $(+--\,-)$ signature of the
Lorentz metric.

In the axiomatic {\it pre\/}metric approach to electrodynamics
proposed recently \cite{gentle,IAS,book}, see also
\cite{Frankel,Post62,Schouten2,Toupin,Truesdell}, the foundations of
electrodynamics are formulated {\it before\/} a metric is introduced.
Thus, different geometrical structures are introduced at different
levels of the construction. This stratification of the structure of
the theory gives a possibility to find out the relations between the
sign assumptions mentioned above.  The aim of this article is to
establish such correlations of signs and even to derive the Lorentz
signature of the underlying metric {}from electric/magnetic reciprocity.

The {\it pre\/}metric approach commences by assuming conservation of
electric charge $\oint{J}=0$ and of magnetic flux $\oint{F}=0$ in
$n$-dimensional spacetime.  In order to attribute to these integral
relations a proper physical meaning as conservation laws, we have to
extract one dimension as `topological time'.  With this topological
assumption, a formulation of Maxwell's theory in $n$ dimensions is
straightforward. Nevertheless, we will restrict the dimensions of the
spacetime that we investigate. Our actual spacetime of 4 dimension is
distinguished {}from other dimensions in that only for $n=4$ the
number of independent components of the electric field $n-1$ equals to
that of the magnetic field $(n-1)(n-2)/2$. In other words, the
electromagnetic field strength $F=(E,B)$ as 2-form in $n$ dimensions
is only a `middle form' for $n=4$. This argument was already mentioned
by Ehrenfest \cite{Ehrenfest}.  A 2-dimensional electron gas in the
context of the quantum Hall effect, for example, can be considered by
meaning of $(1 +2)$-dimensional electrodynamics: Then, for $n=3$, we
have indeed 2 components of the electric field $E$ but only 1
component of the magnetic field $B$.  Such a 3-dimensional model of
electrodynamics is basically different from the electrodynamics in
$n=4$.  {}From now on we will assume $n=4$.

Conventionally, in Maxwell's electrodynamics, a set of assumptions on
signs are postulated partly a priori.  A first assumption is that on
the signs appearing in the Lorentz {\it metric}.  A second one is {\it
  Lenz's rule} which establishes the sign in Faraday's induction law
as opposed to the sign in Amp\`{e}re--Maxwell's somewhat analogous
law.  And a third assumption is the customary set of signs in the {\it
  energy-momentum} tensor of the electromagnetic field. We would like
to disentangle these interrelationships.  It is well-known, for
instance, that Lenz's rule can be derived {}from energy
considerations.  Since these assumptions on signs are postulated in
Maxwell's electrodynamics all together `at the same time', it seems to
be impossible to find out relations between them.  Earlier discussions
of Maxwell's theory with Euclidean signature were given by Zampino
\cite{Zampino} and on the Euclidean Maxwell-Einstein equations by
Brill \cite{Brill}. Both authors didn't use the premetric approach,
even though Brill mentions it.

\section{Premetric electrodynamics and its $(1+3)$-decomposition}
\subsection{Topological structure}
The construction of pre-metric electrodynamics starts by postulating
certain topological conditions on spacetime.

\ax We require spacetime to be a 4-dimensional differential manifold
$X^4$ that allows a smooth foliation of codimension 1. The foliation
is denoted by a smoothly increasing parameter $\sigma\in
(-\infty,\infty)$. Thus we have a partition of $X^4$ into disjoint
path-connected subsets $X^3_\sigma$, which are local homeomorphic to
$\mathbb R^3$.  The parameter $\sigma$ is the prototype of a time
coordinate. We call $\sigma$ the topological time.\eax

\noindent The manifold is considered without metric and without 
connection. The metric is derived in the theory {}from physical
properties of the electromagnetic field.  However, the restrictions on
topology by Axiom 0 are essential for the existence of some physics on
a topological manifold \cite{Geroch}. In particular, they are
necessary for global hyperbolicity and for the existence of a
spinorial structure.

\begin{figure}
\caption{Four-dimensional spacetime and a foliation of codimension 1 
  (see \cite{book}). The coordinate $x^3$ is suppressed.}
\begin{center}
\vspace{2cm}
\includegraphics[width=14cm]{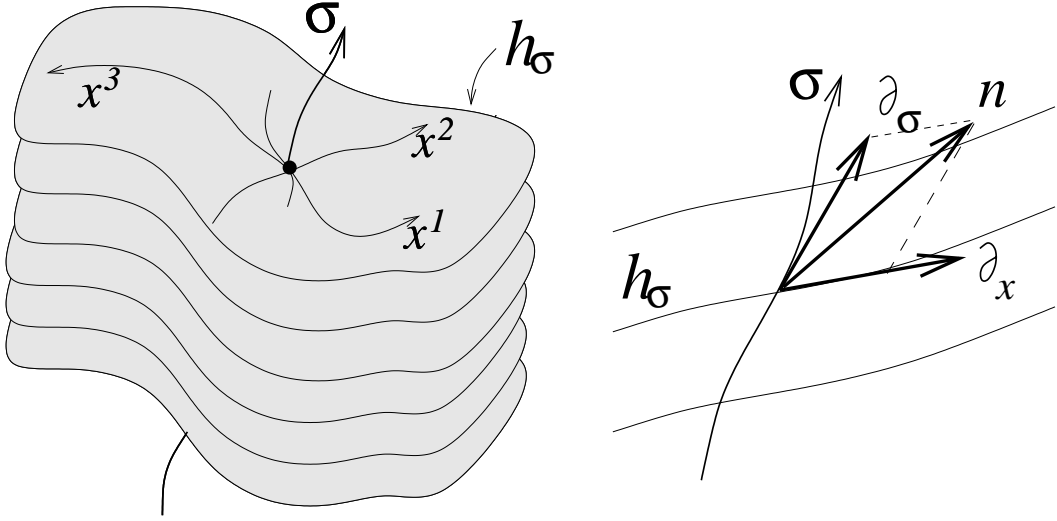}%
\end{center}
\end{figure}
 
With these topological assumptions, we are ready to formulate our
fundamental structures.  The electromagnetic quantities will be
represented by differential forms of different degrees.  For a given
foliation $\sigma$, we are able to decompose them into {\em
  tangential\/} and {\em normal\/} pieces.  An arbitrary $p$-form
$\alpha$ may be decomposed uniquely (for a given $\sigma$) as
\begin{equation}\label{dec0}
\alpha=\beta\wedge d\sigma+\gamma\,,
\end{equation}
where the $(p-1)$-form $\beta$ and the $p$-form $\gamma$ are forms
that lie in the folio, i.e., they satisfy the relations
\begin{equation}\label{dec1}
  e_0\rfloor \beta=e_0\rfloor \gamma=0\,,\qquad e_0={\partial}/
  {\partial \sigma}\,.
\end{equation}
Observe that due to the covariance of the structures used, the forms
$\beta$ and/or $\gamma$ cannot vanish identically. Indeed, if even one
of them is zero for a given foliation $\sigma$, it will be non-vanishing
for a foliation $\sigma'$ that is only `turned' by a small amount.

We define the positive volume element in $X^4$ as
\begin{equation}\label{vol}
{}^{(4)}{\rm vol}=d\sigma\wedge {}^{(3)}{\rm vol}\,,
\end{equation}
where ${}^{(3)}{\rm vol}$ is a positive volume element on a
hypersurface of constant $\sigma$. The 4-dimensional exterior
derivative decomposes as
\begin{equation}\label{dec2}
d=d\sigma\wedge\frac{\partial}{\partial\sigma}+\underline{d}\,,
\end{equation}
where $\underline{d}$ refers to the local coordinates lying in the
hypersurface of constant $\sigma$.  The partial derivative
$\partial/\partial\sigma$ is abbreviated by a dot on top of the
corresponding quantity.

A 3-form $\alpha$ that lies in the folio has specific properties.
Because of antisymmetry, the space-like exterior derivative of such a
form is zero: $\underline{d}\,\alpha=0$.  Moreover, an arbitrary
$\alpha$ is proportional to the volume element ${}^{(3)}{\rm vol}$.
Hence we can distinguish positive and negative 3-forms that are
transversal to the folio.


\subsection{Continuity equation for electric charge}

Let us now postulate the existence of an electric charge current
density.  

\ax The charge current density is a conserved twisted 3-form $J$,
i.e., the 3-form $J$, being integrated over an arbitrary closed 3D
submanifold $C_3\in X^4$, obeys
\begin{equation}\label{charge1}
  \oint\limits_{C_3}J=0\,,\quad \partial C_3=0\qquad \Longrightarrow
  \quad dJ=0\,.
\end{equation}   
\eax The decomposition of $J$ into tangential and normal pieces
relative to the foliation $\sigma$ may be written as
\begin{equation}\label{decJ}
  J=i_{\tt T}\,j\wedge d\sigma+i_{\tt S}\,\rho\,,
\end{equation} 
where we introduced the {\bf{T}}ime and {\bf S}pace factors $i_{\tt
  T},i_{\tt S}$ with values {}from the set $\{+1, -1\}$.  Any factor the
absolute value of which is different {}from $1$ is considered to be
absorbed in the corresponding form. Hence (\ref{decJ}) is the most
general decomposition relative to the given foliation $\sigma$.  We
are going to derive which values of the factors $i_{\tt T},i_{\tt S}$
are in correspondence with the physical interpretation of the forms
$j$ and $\rho$.

We differentiate (\ref{decJ}) and use $\underline{d}\,\rho=0\,$:
\begin{equation}\label{con-x}
   d\,J=i_{\tt T}\,d\, j\wedge d\sigma+i_{\tt S}\,d\,\rho= 
d\sigma\wedge (-i_{\tt T}\,\underline{d}\,j+i_{\tt S}\,\dot{\rho})\,.
\end{equation}
Accordingly, $d\,J=0$ yields
\begin{equation}\label{con}
  \underline{d}\,j-\frac{i_{\tt S}}{i_{\tt T}}\,\dot{\rho}=0\,.
\end{equation}
Let us now take into account the relation between the forms $j$ and
$\rho$.  The 3-form $\rho$ represents the space density of charge,
whereas the 2-form $j$ represents the current density of the {\it
  same} charge.  Then (\ref{con}) has to represent the differential
$(1+3)$-dimensional expression of the conservation law of electric
charge.  Consider the integral version of (\ref{con}).  For this, we
have to choose a compact $3D$ region $\Omega_3$ with the $2D$ boundary
$\partial\Omega_3$ and a normal vector field $n$ directed outwards of
$\partial\Omega_3$, i.e., we assume that the usual conventions are
valid.
\begin{figure}
\caption{Charge conservation in 4-dimensional space (see \cite{book}). 
  Here $t=\sigma$ denotes the topological time.}
\begin{center}
  \vspace{2cm}
  \includegraphics[height=5cm,width=15cm]{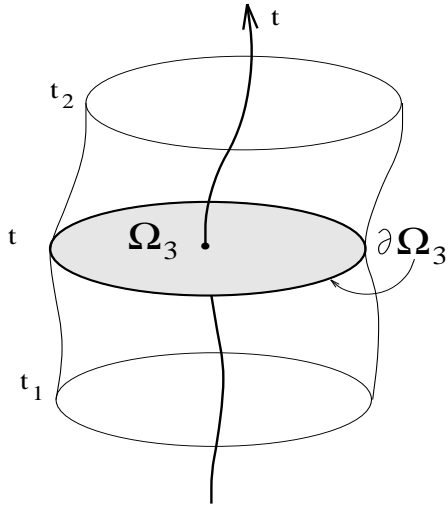}%
\end{center}
\end{figure}
Then, integrating (\ref{con}) over the region $\Omega_3$, 
\begin{equation}\label{con*}
  \int_{\Omega_3}\left( \underline{d}\,j-\frac{i_{\tt S}}{i_{\tt
        T}}\,\dot{\rho}\right)=0\,,
\end{equation}
or by using the divergence theorem,
\begin{equation}\label{con**}
  \int_{\partial\Omega_3}(n\rfloor j)dS=\frac{i_{\tt S}}{i_{\tt
      T}}\,\frac{\partial}{\partial \sigma}\int_{\Omega_3}{\rho}\,.
\end{equation}
The left hand side of (\ref{con**}) represents the flux of charge
through the boundary $\partial\Omega_3$. Let the $\sigma$ axis be
directed in the future. Then, in the case of a positive flux, the
change of the flux in the compact domain bound by $\partial\Omega_3$
has to be negative.

Thus, the right hand side has to represent the change of the charge in
the compact domain bounded by $\partial\Omega_3$.  Consequently,
\begin{equation}\label{cur}
  i_{\tt S}=-i_{\tt T}\,. 
\end{equation}
In this case, eq.(\ref{con}) turn out to be the continuity equation
\begin{equation}\label{con***}
  \underline{d}\,j+\dot{\rho}=0\,,
\end{equation}
which represents the appropriate conservation law. 

A pointwise charge density, for example, 
\begin{equation}\label{point-1}
  \rho=e\,\delta^{(3)}(\vec{r}-\vec{r}_0)\,dx\wedge dy\wedge dz\,,
\end{equation}
provided we define the current in the conventional way by 
\begin{equation}\label{point-2}
j=v\rfloor \rho\,,
\end{equation}
fulfills (\ref{con}) only for the parameters (\ref{cur}).
Indeed, differentiating (\ref{point-1}), we obtain
\begin{equation}\label{point-3}
  \frac{\partial\rho}{\partial \sigma}=\frac{\partial \rho}{\partial
    x^i_0} \frac{\partial x^i_0}{\partial \sigma}=-\frac{\partial
    \rho}{\partial x^i}v^i=-d(v\rfloor \rho)=-dj\,.
\end{equation}
Consequently, by (\ref{cur}), the true decomposition of the
4-dimensional current $J$, with the convention (\ref{point-2}), reads
\begin{equation}\label{cur*}
  J=i_{\tt T}\left(j\wedge d\sigma -\rho \right)\,. 
\end{equation}
The ``$-$'' sign in this decomposition originates, via (\ref{con-x}),
{}from the odd degree of the form $\underline{d}\,j$.  Nevertheless, the
continuity equation (\ref{con***}) involves the ``$+$'' sign.  This
decomposition can be also considered as a choice of the proper
direction of the $\sigma$-axis.

\subsection{Inhomogeneous Maxwell equation}

By de Rham's theorem, the inhomogeneous Maxwell equation is a
consequence of the charge conservation law (\ref{charge1})${}_1$,
\begin{equation}\label{ax1a}
J=dH\,,
\end{equation} 
where $H$ is the twisted 2-form of the {\it electromagnetic
  excitation}, which has the absolute dimension of charge. 
In the $(1+3)$-decomposition of the excitation
\begin{equation}\label{decH}
H=h_{\tt T}\,{\mathcal H}\wedge d\sigma+h_{\tt S}\,{\mathcal D}\,,
\end{equation}
we introduced again the sign factors $h_{\tt T}$ and $h_{\tt S}$ with
values {}from the set $\{+1, -1\}$.  We differentiate (\ref{decH})
\begin{eqnarray}\label{inhomMax-0}
  dH&=&h_{\tt T}\,\underline{d}{\mathcal H}\wedge d\sigma+h_{\tt
    S}\,\underline{d}{\mathcal D}+ d\sigma\wedge\dot{\mathcal D}\nonumber\\ 
  &=&d\sigma\wedge \left(h_{\tt T}\,\underline{d}\,{\mathcal H}+{h_{\tt
        S}}\,\dot{\mathcal D}\right) +h_{\tt S}\,\underline{d}{\mathcal D}\,.
\end{eqnarray}
Thus the inhomogeneous Maxwell equation (\ref{ax1a}), with the source
(\ref{cur*}), decomposes according to
\begin{equation}\label{inhomMax}
  {h_{\tt T}\,\underline{d}\,{\mathcal H}+{h_{\tt S}}\,\dot{\mathcal D} =
    i_{\tt T}j\,,\qquad\quad h_{\tt S}\,\underline{d}\,\mathcal D}=-i_{\tt
    T}\rho\,.\end{equation}

\subsection{Lorentz force}

This axiom introduces implicitly the {\it electromagnetic field
  strength} $F$ as an independent concept via the mechanical concept
of force and the existence of prescribed electric test currents.  

\ax The Lorentz force density (a twisted covector-valued 4-form) is
postulated as
\begin{equation}\label{axiom2}
  f_\alpha= (e_\alpha\rfloor F) \wedge J\,,
\end{equation} 
where $e_\alpha$ is the frame and $F$ an untwisted 2-form.  \eax
The field strength $F$ has the absolute dimension of magnetic flux.
Its $(1+3)$-decomposition reads
\begin{equation}\label{axiom2b}
F=f_{\tt T}\,E\wedge d\sigma+f_{\tt S}\,B\,,
\end{equation} 
where the sign factors $f_{\tt T}$ and $f_{\tt S}$ take values {}from
$\{+1, -1\}$.

Thus, the Lorentz force density (\ref{axiom2}) decomposes
according to
\begin{equation}\label{lor5}
  f_0=-i_{\tt T}f_{\tt T}\,E\wedge j\wedge d\sigma\,
\end{equation}
and
\begin{equation}\label{lor6}
  f_\mu=\bigg[i_{\tt T}f_{\tt T}(e_\mu\rfloor E)\rho+
i_{\tt T}f_{\tt S}(e_\mu\rfloor B)\wedge j\bigg]\wedge d\sigma\,, \qquad \mu=1,2,3\,.
\end{equation}
\subsection{Homogeneous Maxwell equation}

{}From a formal point of view, we have four equations (\ref{axiom2})
for six components of the 2-form $F$. Additional condition have to be
employed in order to fix them uniquely.  \ax The conservation of
magnetic flux is postulated; as a consequence, we find the homogeneous
Maxwell equation,
\begin{equation}\label{axiom3}
  \oint\limits_{C_2}F=0\,,\quad \partial C_2=0\,,\qquad
  \Longrightarrow \quad dF=0\,,
\end{equation} 
for any closed submanifold $C_2$. \eax By de Rham's theorem, 
(\ref{axiom3})${}_1$ yields
\begin{equation}\label{axiom3b}
F=dA\,,
\end{equation}
where $A$ is the untwisted 1-form of {\it the electromagnetic
  potential}.  According to (\ref{axiom3}), the field strength $F$
would be only determined up to the differential of an arbitrary
1-form. However, the expression for the Lorentz force (\ref{axiom2})
defines it uniquely. In accordance with (\ref{axiom2b}), the
homogeneous Maxwell equation $dF=0$ decomposes as
\begin{eqnarray}\label{homMax-1}
  dF&=&f_{\tt T}\,\underline{d}\,E\wedge d\sigma+ f_{\tt
    S}\left(\underline{d}\,B +d\sigma \wedge
    \dot{B}\right)\nonumber\\&=& d\sigma\wedge \left( f_{\tt
      T}\,\underline{d}\,E+f_{\tt S}\,\dot{B}\right)+f_{\tt
    S}\,\underline{d}\,B=0\,.
\end{eqnarray}
Thus, 
\begin{equation}\label{homMax}
  f_{\tt T}\,\underline{d}\,E+f_{\tt S}\dot{B} = 0\,,\qquad\quad
  \underline{d}\,B=0\,.
\end{equation} 
\section{Energy-momentum current}
\subsection{Energy-momentum current and Lorentz force}

Up to now, we introduced the electromagnetic field $(H,F)$ and its
field equations $dH\!=\!J$ and $dF\!=\!0$. At this stage of our
construction, no specific geometric structure is provided. We still
deal with a (topological) differential manifold without metric and
without connection. The energy-momentum content of the electromagnetic
field is described by means of a covector valued 3-form. We will
denote it by $\Sigma_\alpha$, with $ \alpha=0,1,2,3$.  It should be
constructed in a covariant way in terms of the fields $H$ and $F$ and
the frame field $e_\alpha$.  Certainly, for a system involving the
electromagnetic field together with a source $J$, the current
$\Sigma_\alpha$ of the electromagnetic field alone cannot be
conserved.  Moreover, the Lorentz force density $f_\alpha$ has to be
treated as a {\it source\/} of the energy-momentum current, or,
conversely, $\Sigma_\alpha$ is a kind of generalized {\it potential\/}
for $f_\alpha$. Since $J\!=\!dH$, the Lorentz force contains the
derivative of the field $H$. For the energy-momentum current we assume
that it depends only on the fields, but not on their derivatives.

Consider the 4-form $d\Sigma_\alpha$.  Due to the linearity of the
exterior derivative operator, this quantity can be represented as a
sum of two terms: $d\Sigma_\alpha={\mathfrak f}_\a+X_{\alpha}$. In the
first term ${\mathfrak f}_\a$ the derivatives of the electromagnetic
field are involved, whereas in the second term $X_\alpha$ there occur
the derivatives of the frame field.  The first term describes how the
energy-momentum current changes under a temporal and spatial variation
of the electromagnetic field. This is exactly what the Lorentz force
density is supposed to mean. Thus, we have to assume ${\mathfrak
  f}_\a=f_\alpha$ and, consequently,
\begin{equation}\label{new1}
  d\Sigma_\alpha-f_\alpha=X_\a\,.
\end{equation}
Recall that the term $X_\a$ in (\ref{new1}) does {\em not\/} involve
derivatives of $F$ and $H$.

We substitute the expression for the Lorentz force
$f_\alpha=(e_\a\rfloor F)\wedge J$ in (\ref{new1}) and use the
inhomogeneous Maxwell equation $J=dH$. Moreover, we may add an
arbitrary term proportional to the left-hand-side of the homogeneous
Maxwell equation $dF=0$:
\begin{equation}\label{em0a}
  d\Sigma_\alpha= (e_\a\rfloor F)\wedge dH+c\, (e_\a\rfloor H)\wedge
  dF+X_\a\,.
\end{equation}
The right hand side already involves the exterior differential
operator $d$.  Thus, the current $\Sigma_\alpha$ has to be linear in
$H$ and $F$ and no derivatives can occur.  Accordingly, we are led to
the expression
\begin{equation}\label{em1a}
  \Sigma_\alpha= a(e_\a\rfloor F)\wedge H+b (e_\a\rfloor H)\wedge F\,,
\end{equation}
with undefined constants $a$ and $b$. 

Thus we conclude that the energy-momentum current $\Sigma_\alpha$ has
to be described by a covector-valued 3-form bilinear in the fields $H$
and $F$ and that it is physically meaningful only for definite values
of the constants $a$ and $b$.  Moreover, the values of the constants
$a$ and $b$ have to guarantee the compatibility with the first four
axioms. As a covector-valued 3-form, $\Sigma_\a$ has in general
$4\times 4=16$ independent components.

Although (\ref{new1}) is covariant under local frame transformations,
its left and right hand sides, respectively, are certainly not
covariant themselves.  However, this separation is meaningful in the
following sense: Due to the linearity of the derivative operator $d$
(Leibniz rule), an arbitrary local change of the frame only yields
additional terms that are proportional to the fields $H$ and $F$ but
not to their derivatives $dH$ and $dF$, see \cite{Itin-2003} for
details.  Indeed, under a linear transformation of the frame,
$e_{\alpha'}=L_{\alpha'}{}^\beta e_\beta$, the (covariant)
energy-momentum current transforms as $\Sigma_{\alpha'}=
L_{\alpha'}{}^\beta\Sigma_\beta$ and the Lorentz force as
$f_{\alpha'}= L_{\alpha'}{}^\beta f_\beta$.  We multiply (\ref{new1})
by $L_{\alpha'}{}^\alpha$. Then
\begin{equation}\label{new1x}
  d\left(L_{\alpha'}{}^\beta\Sigma_\beta\right)-d L_{\alpha'}{}
  ^\beta\wedge \Sigma_\beta-L_{\alpha'}{}^\beta
  f_\beta=L_{\alpha'}{}^\beta X_\beta
\end{equation}
or
\begin{equation}\label{new1xx}
  d\Sigma_{\alpha'}- f_{\alpha'}=L_{\alpha'}{}^\beta X_\beta +d
  L_{\alpha'}{}^\beta\wedge \Sigma_\beta=:X_{\alpha'}\,.
\end{equation}
Thus, although the separate terms in (\ref{new1}) are not covariant,
the property of the left hand side to contain the derivatives of the
electromagnetic field $(F,H)$ and of the right hand side to contain
the derivative of the frame field $e_\alpha$ is preserved under an
arbitrary local linear transformation of the frame.

Accordingly, in (\ref{new1}), $dH$ and $dF$ appear only on the left
hand side and this is preserved under local linear frame
transformations.  Consequently, the numerical values of the
coefficients $a$ and $b$ of the energy-momentum current (\ref{em1a})
have to be chosen such that in (\ref{new1}) the derivative terms $dH$
and $dF$ on its left hand side cancel each other.
Substituting (\ref{em1a}) in (\ref{new1}) we obtain
 \begin{equation}\label{em4a}
   a\, d(e_\a\rfloor F)\wedge H-a(e_\a\rfloor F)\wedge dH+b\,
   d(e_\a\rfloor H)\wedge F-f_\alpha=X_\a\,
\end{equation}
or, equivalently,
 \begin{equation}\label{em5a}
   (a\, \L_\a F\wedge H+b\, \L_\a H\wedge F)-(a-b+1)f_\alpha=X_\a\,,
\end{equation}
where we introduced a short-hand notation for the Lie derivative operator 
$\L_\a:=\L_{e_\a}=d(e_\a\rfloor)+e_\a\rfloor d$. 

The terms with $dH$ and $dF$ in the first two terms of (\ref{em5a})
have to vanish independently of the third term. Indeed, they have to
vanish even in the sourcefree case when the Lorentz force is zero.
Hence,
 \begin{equation}\label{em6a}
a-b+1=0\,.
\end{equation}
Eq.(\ref{em5a}) now simplifies to
\begin{equation}\label{em7a}
  a\, \L_\a F\wedge H+b\, \L_\a H\wedge F=X_\a\,.
\end{equation}

So far, the fields $H$ and $F$ are completely independent.  The
derivatives on the left hand side cannot cancel one another for any
nonzero values of the parameters $a$ and $b$.  This situation is not
surprising.  Indeed, the electromagnetic field $(H,F)$ has 12
independent components, whereas it is governed by the 8 independent
field equations (\ref{ax1a}) and (\ref{axiom3}). Thus the system is
underdetermined and it is natural that it does not provide a uniquely
defined energy-momentum current.  We postpone the treatment of
(\ref{em7a}) to the next section where a relation between the fields
$H$ and $F$ will be introduced.

Our intermediate result is then, see (\ref{em6a}), that the
energy-momentum current reads
\begin{equation}\label{intermediary}
  \Sigma_\alpha= a(e_\a\rfloor F)\wedge H+(a+1) (e_\a\rfloor H)\wedge
  F\,.
\end{equation}
If we transvect (\ref{intermediary}) with the coframe $\vartheta^\a$,
we find the so-called {\it trace\/} of the energy-momentum current
\begin{equation}\label{em1aaa}
  \vartheta^\alpha\wedge\Sigma_\alpha=2 (2a+1) F\wedge H\,.
\end{equation}
This expression is a 4-form and has 1 independent component. It is
equivalent to an irreducible piece of $\Sigma_\a$. The tracefree piece of
the energy-momentum current with 15 independent components can be
defined as
\def\negenspace{\kern-1.1em}
\def\quer{\negenspace\nearrow}
\def\negenspaceexp{\kern-0.5em}
\def\querexp{\negenspaceexp\nearrow}
\begin{equation}\label{tracefree}
\Sigma\quer_{\alpha}:=\Sigma_\a-\frac14\,e_\a\rfloor
  \left(\vartheta^\b\wedge\Sigma_\b\right)\,.
\end{equation}

\subsection{Axiom on energy-momentum}

{}From field theory we know, see \cite{Lopu}, that for a non-scalar
field the trace of the energy-momentum tensor is always related to the
mass of the corresponding field.  Thus, in pre-metric electrodynamics,
the electromagnetic field (the ``photon'' field) is massless provided
$\vartheta^\b\wedge\Sigma_\b$ vanishes or, equivalently, $a=-1/2$.
This relation will be justified in an alternative way (without
referring to the trace) in the next section. Still, we can already now
formulate our axiom:

\ax The energy-momentum current of the electromagnetic field is the
covector-valued 3-form
 \begin{equation}\label{em1aa}
   \Sigma_\alpha:= \frac 12\left[ (e_\a\rfloor H)\wedge F-
     (e_\a\rfloor F)\wedge H\right]\,.
\end{equation}
\eax 

\noindent Accordingly, $\vartheta^\a\wedge\Sigma_\a=0$, i.e., our 
energy-momentum current is traceless
$\Sigma\quer_{\alpha}=\Sigma_{\alpha}$ and carries 15 independent
components.


Let us now decompose this current into time and space components.
Since we are interested in the {\em energy} of the electromagnetic
field, it is sufficient to discuss the ``time'' ($t=\sigma$) component
$\Sigma_0$.  Because of $e_0\rfloor d\sigma=1$ and
$e_0\rfloor{\mathcal H}=e_0\rfloor{\mathcal D}=e_0\rfloor E=e_0\rfloor
B=0$ (forms lie in the folio $\sigma$), we find $e_0\rfloor
H=\,-h_{\tt T}\,{\mathcal H}$ and $e_0\rfloor F=-f_{\tt T}\, E$.
Thus, the decomposition of $\Sigma_0$ turns out to be
\begin{eqnarray}\label{em2aa}
  \Sigma_0&=&\frac 12 f_{\tt T}E\wedge (h_{\tt T}{\mathcal H}\wedge
  d\sigma+h_{\tt S}{\mathcal D}) - \frac 12 h_{\tt T}{\mathcal H}\wedge
  (f_{\tt T}E\wedge d\sigma+ f_{\tt S}\, B ) \nonumber\\ &=& h_{\tt
    T}f_{\tt T}E\wedge{\mathcal H}\wedge d\sigma+\frac 12 h_{\tt S}f_{\tt
    T}\, {\mathcal D}\wedge E -\frac 12h_{\tt T}f_{\tt S}\,{\mathcal H}\wedge
  B\,.
\end{eqnarray}
The relation between the sign of the Lorentz force and the sign of the
energy-momentum current can be read off from
\begin{equation}    d\Sigma_0 = f_0 + X_0\,, \end{equation}
see (\ref{new1}).

\subsection{Reciprocity symmetry}

In the absence of a source $J$, the Maxwell field equations $dH=0$ and
$dF=0$ do not change under the transformation $H\rightarrow \mu F$ and
$F\rightarrow \nu H$. Here $\mu$ and $\nu$ are non-vanishing
dimensionful {\em constants:\/} $\mu\ne 0$ and $\nu\ne 0$.  This
symmetry is expected to be preserved in the energy-momentum current
$\Sigma_\a$ of the electromagnetic field since this current has to be
independent of the source. A look at (\ref{em1aa}) shows that this
transformation yields
\begin{equation}\label{em4aaa}
  \Sigma_\a \rightarrow\frac{\mu\nu}{2}\,\left[-\left(e_\a\rfloor
      H\right)\wedge F + \left(e_\a\rfloor F\right)\wedge H
  \right]=-\mu\nu\Sigma_\a \,.
\end{equation}
Consequently, 
\begin{equation}\label{em4aax}
\mu\nu=-1
\end{equation}
yields the invariance $\Sigma_\a\rightarrow  \Sigma_\a$. 
Hence $\Sigma_\a$ is only invariant under the
transformation $H\rightarrow \mu F\,,\,F\rightarrow -H/\mu$. 

In this analysis of $\Sigma_\alpha$ we used the symmetry
$H\rightarrow \mu F$ and $F\rightarrow -H/\mu$ of the sourcefree
Maxwell equations. However, as a rule, one should preferably
investigate a symmetry in Lagrangians or Hamiltonians (i.e.,
energy-momentum expressions), see a corresponding remark of
Staruszkiewicz \cite{Star}. Therefore we sharpen our notions and
recognize that the energy-momentum current (\ref{em1aa}) (not,
however, the Maxwell equations) is invariant under the transformation
\begin{equation}\label{em7aa}
H\to  \zeta F\,, \quad F\to -\frac 1\zeta\, H\,.
\end{equation}
We refer to it as {\it electric/magnetic reciprocity}.  The function
$\zeta=\zeta(x)$ is an arbitrary twisted 0-form (pseudo-scalar
function) of dimension $[\zeta]=[H]/[F]=1/$resistance.  Because of
$H=h_{\tt T}{\mathcal H}\wedge d\sigma+h_{\tt S}{\mathcal D}$ and $F=f_{\tt
  T}E\wedge d\sigma +f_{\tt S}B$, the $(1 + 3)$-decomposition of the
reciprocity transformation (\ref{em7aa}) reads
\begin{equation}\label{duality1aa} \hspace{-30pt}
  \left\{ {\mathcal H}\rightarrow \zeta \frac {f_{\tt T}}{h_{\tt
        T}}E\,,\quad {\mathcal D} \rightarrow \zeta \frac {f_{\tt
        S}}{h_{\tt S}} B \right\} \,,\qquad \left\{E\rightarrow
    \,-\frac 1\zeta\, \frac {h_{\tt T}}{f_{\tt T}}{\mathcal H}\,,\quad
    B\rightarrow - \frac 1\zeta \frac {h_{\tt S}}{f_{\tt S}}{\mathcal
      D}\right\}\,.
\end{equation} 
Accordingly, electric quantities are replaced by magnetic ones and
vice versa.

\section{Spacetime relation}

\subsection{Constitutive tensor of spacetime}

So far, the electromagnetic field $(H,F)$ is underdetermined. It has
12 independent components that satisfy only 8 independent equations
$dH=J$ and $dF=0$.  We need a {\it spacetime relation\/} linking the
excitation $H$ to the field strength $F$.  First of all, we require
that this ``constitutive law for vacuum'' is {\em local}, that is, $H$ at
a certain event with coordinates $x=(x^0,x^1,x^2,x^3)$ depends only on
$F$ at the same event. Neither differentials nor integrals are allowed.
With a local operator $\kappa=\kappa(x)$ we can write $H=\kappa (F)$.
Here $\kappa$ is a twisted operator that does not depend on $H$ and
$F$. Furthermore, we require {\em linear\/}ity of this operator. With
arbitrary constants $a$ and $b$, we have
\begin{equation}\label{sr1}
  H=\kappa (F)\,, \qquad \kappa(a\,\Phi+b\,\Psi)=a\,\kappa
  (\Phi)+b\,\kappa(\Psi)\,,
\end{equation}
for arbitrary 2-forms $\Phi$ and $\Psi$.

We decompose $H$ and $F$ into their components
\begin{equation}\label{sr2}
H=\frac 12 H_{\a\b}\,\vartheta^\a\wedge\vartheta^\b\,,\quad
F=\frac 12 F_{\a\b}\,\vartheta^\a\wedge\vartheta^\b\,.
\end{equation}
Here $\vartheta^\a$ denotes the coframe. Using linearity, we can
represent the linear operator $\kappa$ componentwise:
\begin{equation}\label{sr3x}
  H=\kappa (F)\quad \Longleftrightarrow\quad H_{\alpha\beta}=\frac 12
  {\kappa_{\alpha\beta}}^{\gamma\delta}F_{\gamma\delta}\,.
\end{equation}
For a compact representation of the {\it constitutive tensor\/}
${\kappa_{\alpha\beta}}^{\gamma\delta}$ of spacetime (of the
``vacuum''), it is useful to apply a formalism with indices running
{}from 1 to 6: $I,J,\dots=\{1,2,3,4,5, 6\}=\{01,02,03,23,31,12\}$.
The constitutive tensor takes now the form of a 6D-tensor
${\kappa_I}^J$ with 36 independent components. Two other important
quantities, which also possess a $6\times 6$ representation, are the
Levi-Civita symbols $\epsilon^{IJ}$ and $\epsilon_{IJ}$.  These
symmetric tensor densities play the role of a quasi-metric in the
6-dimensional vector space of 2-forms.  In particular, they can be
used for the raising and lowering of the indices of a tensor.
Nevertheless, this procedure has to be carefully distinguished {}from
the usual raising and lowering of indices by means of a 4D metric.
After all, we are dealing here with {\em pre\/}metric electrodynamics.

In this way we define the tensor density
 \begin{equation}\label{sr4}
   \chi^{IJ}:=\epsilon^{IM}{\kappa_M}^J\,, \qquad
   {\kappa_I}^J=\epsilon_{IM}\,\chi^{MJ}\,.
\end{equation}
The constitutive tensor, in the form $\chi^{IJ}$ as $6\times 6$
matrix, can be straightforwardly decomposed into its irreducible
parts, the traceless symmetric part, the antisymmetric part, and the
trace:
 \begin{equation}\label{sr5}
\chi^{IJ}={}^{(1)}\chi^{IJ}+{}^{(2)}\chi^{IJ}+{}^{(3)}\chi^{IJ}\,.
\end{equation}
The {\it principal\/} part ${}^{(1)}\chi^{IJ}$ has 20 independent
components, the {\it skewon\/} part ${}^{(2)}\chi^{IJ}$ 15 components,
and the {\it axion\/} part ${}^{(3)}\chi^{IJ}=\epsilon^{IJ}\alpha$ is
equivalent to one pseudo-scalar field $\alpha$.

\subsection{Energy-momentum current once more}

Now we are able to go back to (\ref{em7a}):
\begin{equation}\label{em7ax}
  a\, \L_\a F\wedge H+b\, \L_\a H\wedge F=X_\a\,.
\end{equation} 
We assume the local and linear spacetime relation (\ref{sr3x}). We use
the linearity of the Lie derivative and find
\begin{equation}\label{em11a}
\L_\a H\wedge F=\L_\a \kappa (F)\wedge F=\kappa (\L_\a F)\wedge F+Y_\a\,.
\end{equation}
Here $Y_\a$ denotes those terms that are obtained by taking the Lie derivative 
of the operator $\kappa$. Thus $Y_\a$ does not involve $dH$ and $dF$.

In order to proceed, we need an another property of $\kappa$. We assume
that in (\ref{sr5}) the skewon piece vanishes. Then $\chi$ is {\em
  symmetric} and
\begin{equation}\label{symm}
\kappa(\Phi)\wedge\Psi=\Phi\wedge\kappa(\Psi)\,,
\end{equation}
for any 2-forms $\Phi$ and $\Psi$.  Since $\L_\a F$ is a 2-form,
(\ref{em11a}) can be rewritten as
\begin{equation}\label{em12a}
  \L_\a H\wedge F=\L_\a F\wedge H+Y_\a\,.
\end{equation}
We substitute this in (\ref{em7ax}) and find
 \begin{equation}\label{em13a}
(a+b)\, \L_\a F\wedge H=X_\a-bY_\a\,.
\end{equation}

Recall that only in the left hand side of this equation the derivative
of the field $F$ is involved.  Thus, in addition to (\ref{em6a}), we
have a second relation between the coefficients, namely
\begin{equation}\label{new2}
  a+b=0\,.
\end{equation}
Consequently,
 \begin{equation}\label{em14a}
a=-\frac 12\,, \qquad b=\frac 12\,,
\end{equation}
and, finally, we recover the energy-momentum current of Axiom 4.
Observe that in our analysis we applied Maxwell's field equations and
the expression for the Lorentz force, that is, Axioms 1, 2, and 3, as
well as {\em locality, linearity, and symmetry\/} of the spacetime
relation. However, neither a specific metric nor a connection have
been used.

A word of caution is in order: Our Axiom 5 is logically independent of
the spacetime relation. As soon as $H$ and $F$ are specified, an
energy-momentum current \`a la Axiom 5 always exists. If we assume
additionally a local, linear, and symmetric spacetime relation, then
$f_\alpha=d\Sigma_\alpha+($terms depending only on H and F$)$.  Thus a
specific spacetime relation implies a specific form of the equation
that is related to energy-momentum conservation.

\subsection{Electric/magnetic reciprocity of the spacetime relation}
\label{emr}

Electric/magnetic reciprocity means that a specific exchange of the
fields $H$ and $F$ preserves the energy-momentum current
$\Sigma_\alpha$. This is possible since $\Sigma_\alpha$ is
algebraically expressed in terms of the fields $H$ and $F$.
Accordingly, it is natural to assume that electric/magnetic
reciprocity is also applicable to another algebraic equation --- the
spacetime relation (\ref{sr3x}).  Hence, with (\ref{em7aa}) and
(\ref{sr1})$_2$, we obtain
 \begin{equation}\label{sr6}
   H=\kappa(F) \qquad \Longrightarrow \qquad \zeta \,F=-\frac 1\zeta\,
   \kappa(H)= -\frac 1\zeta\, \kappa^2(F)\,.
\end{equation}

Since $F$ is arbitrary, $\kappa^2$ turns out to be proportional to the
identity operator of the 6D vector space:
 \begin{equation}\label{sr7}
\kappa^2=-\zeta^2\,{\mathbb {I}}_6\,.
\end{equation}
The energy-momentum current was electric/magnetic reciprocal for
arbitrary $\zeta$. Not so for the spacetime relation. The components
of the linear operator $\kappa$ are directly observable. Thus
$\kappa^2$ must not depend on an arbitrary function $\zeta^2$. For
this reason, we take the trace of (\ref{sr7}) and resolve it with
respect to $\zeta^2$:
\begin{equation}\label{zetasquare}
  \zeta^2 = -\,{\frac 1 {6}}\,{\rm Tr}({\kappa}^2) = -\,{\frac 1
    {6}}\,{\kappa}_{I}
  {}^{K}\,{\kappa}_{K}{}^{M}\,.
\end{equation}
Accordingly, $\zeta$ is no longer an arbitrary function, it is rather
expressed in terms of the constitutive tensor $\kappa$.

The square of a real {\em pseudo\/}-scalar field is a real scalar
field. Hence, instead of the pseudo-scalar field $\zeta$, we may
introduce a ``true'' scalar field $\lambda$ by means of the relation
$\lambda^2:=\zeta^2$.  The physical dimension of this new scalar field
is $[\lambda]=1/$resistance.  Accordingly,
 \begin{equation}\label{sr8x}
   \kappa^2=-\lambda^2\,{\mathbb {I}}_6\,.
\end{equation}
Therefore, we are able to introduce an almost complex structure on the
6D vector space
\begin{equation}\label{sr9}
  {\mathbb J}:=\frac 1\lambda\, \kappa\,, \qquad {\mathbb J}=\sqrt
  {-{\mathbb I}_6}\,.
\end{equation}

\subsection{Signature emerges}

The principal part of the constitutive tensor obeys the symmetry
$\chi^{IJ}=\chi^{JI}$.  Let us put the other two other parts to zero,
namely the skewon and the axion parts.  Consequently, for arbitrary
2-forms $\Phi$ and $\Psi$, we have the symmetry
\begin{equation}\label{sr13}
\kappa(\Phi)\wedge \Psi=\kappa(\Psi)\wedge \Phi\,, \qquad 
{\mathbb J}(\Phi)\wedge \Psi={\mathbb J}(\Psi)\wedge \Phi\,.
\end{equation}
Moreover, 
\begin{equation}\label{sr13a}
{\mathbb J}^2 \Phi=-\Phi\,.
\end{equation}
In such a way we have constructed a local and linear operator
${\mathbb J}$ that is (i) symmetric, (ii) maps twisted 2-forms to
untwisted ones, and (iii), if squared, equals to the negative of the
identity operator. These properties, here found for an operator acting
on 2-forms, are the those of the Hodge star operator.  Therefore, our
operator ${\mathbb J}$ corresponds to the Hodge star operator
$*_{(g)}$, constructed (uniquely) {}from some metric $g$ given on the
manifold.  The square of the Hodge operator, acting on a $p$-form
$\omega$, is given by
\begin{equation}\label{sr14}
  *^2\,\omega =(-1)^{p(n-p)+{\rm ind}}\,\omega\,,
\end{equation}
where $n$ is the dimension of the manifold and ${ind}$ the index of
the metric (the number of the minus signs in the signature).  For
2-forms $\Phi$ on a $4D$-manifold, this formula yields
\begin{equation}\label{sr14a}
*^2\Phi =(-1)^{\rm ind}\,\Phi\,.
\end{equation}
Comparing this relation with (\ref{sr13a}) we derive 
\begin{equation}\label{sr14b}
{\rm ind}=1,\,3\,.
\end{equation}
The unique signature of such indices on a $4D$-manifold is Lorentzian
$(-1,+1,$ $+1,+1)$ or, equivalently, $(+1,-1,-1,-1)$.

Thus we obtain the important result: {\it The electric/magnetic
  reciprocity of the energy-momentum current, if applied to a local,
  linear, and symmetric spacetime relation, yields an operator that is
  equivalent to the Hodge operator of a metric with Lorentzian
  signature.} Accordingly, the answer to the question posed in the
title of our paper is clearly affirmative.

\subsection{Maxwell-Lorentz spacetime relation}

So far, we considered electrodynamics (field equations and
conservation laws) on a metric-free background.  We have shown that a
metric of Lorentzian signature is singled out by its correspondence to
a specific symmetry requirement of the spacetime relation, namely to
its electric/magnetic reciprocity (see Sec.\ref{emr}).  Let us
consider now a manifold endowed with a metric $g$ of a certain
signature.  Our goal is twofold. On the one hand, we want to establish
which of the sign factors of electrodynamics is induced by the
Lorentzian signature.  On the other hand, we want to examine which
sign factors and, correspondingly, which laws of electrodynamics
emerge in the case of a Euclidean metric. The last question was
discussed by Zampino \cite{Zampino} and Brill \cite{Brill}. Since sign
factors as well as the signature of the metric do not depend on a
point, it is enough to deal with a local metric $g$ referred to
orthogonal axes.  For our foliated manifold it means that we should
take orthonormal frames in a folio and a $\sigma$-axis normal to the
folio.  Thus, the components of the metric read
\begin{equation}\label{lr0}
  g_{\a\b}={\rm diag}\bigg((-1)^{s_0}\,c^2,(-1)^{s_1},+1,+1\bigg)\,,
  \qquad s_0,s_1\in \{0,1\}\,,
\end{equation}
with $c$ as velocity of light.  This expression for the components of
the metric embodies all possible signatures, see Table 1.
\begin{table}[htbp]
  Table 1. Signature and the exponents $s_0,\,s_1$ in Eq.(\ref{lr0})
  \medskip
\begin{center}
{\begin{tabular}{||c||c|c||}
\hline\hline
Signature&$s_0$&$s_1$\\ \hline\hline
Minkowski (aka Lorentz) & 1 & 0  \\ \hline
Euclid & 0 & 0 \\ \hline
$(-,-,+,+)$ & 1 & 1 \\ \hline\hline
\end{tabular}}
\end{center}
\end{table}

We specialize the constitutive tensor by demanding vanishing of the
skewon and axion pieces. These two additional fields to ordinary
electrodynamics do not affect the sign factors.
The spacetime relation takes now the standard form 
\begin{equation}\label{lr1}
  H=\lambda *F\,, 
\end{equation}
wherein the Hodge operator $*$ is defined in terms of the metric
(\ref{lr0}). The scalar function $\lambda$, a dilaton type field, can
also be considered as an addendum to standard electrodynamics.  By
requiring $\lambda$ to be constant, we discard also this field.

For the $(1+3)$ decomposition of (\ref{lr1}) we introduce the
3-dimensional Hodge operator ${\underline {*}}$.  For the diagonal
metric (\ref{lr0}), the interrelation between $*$ and ${\underline *}$
takes a rather simple form,
\begin{equation}\label{lr2}
  *(d\sigma\wedge E)=\,(-1)^{s_0}\,\frac{1}{c}\, {\underline *}E\,, \qquad
  *B=c\,d\sigma\wedge\, {\underline *}B\,,
\end{equation}
see \cite{book}. Because of (\ref{sr14}), we have ${\underline
  *}^2=(-1)^{s_1}$. Thus the $(1+3)$ decomposition of (\ref{lr1}),
with (\ref{decH}), (\ref{axiom2b}), and (\ref{lr2}), reads
\begin{equation}\label{lr3}
  h_{\tt T}\,{\mathcal H}\wedge d\sigma+h_{\tt S}\,{\mathcal D}=
  -\lambda \left[\,(-1)^{s_0}f_{\tt T}\,\frac{1}{c}\;{\underline *}\,E-
    f_{\tt S}\,c\,d\sigma\wedge\, {\underline *}\,B\right]\,
\end{equation}
or, equivalently, 
\begin{equation}\label{lr4}
  E=\frac c\lambda\,(-1)^{s_0+s_1+1}\,\frac {h_{\tt S}}{f_{\tt T}}\;
  {\underline *}\,{\mathcal D}\,,\qquad B=\frac{ 1}{\lambda
    c}\,(-1)^{s_1+1}\,\frac{h_{\tt T}}{f_{\tt S}}\; {\underline
    *}\,{\mathcal H}\,.
\end{equation}
Hence instead of the four 3-dimensional forms $E, {\mathcal D}, {\mathcal
  H},B$, we can now consider only one pair of forms.  We choose the
excitation $H=({\mathcal H},{\mathcal D})$ since $H$ is straightforwardly
determined by its sources $\rho$ and $j$.

\section{Positivity of the electromagnetic energy}

Let us now rewrite the ``time'' ($t=\sigma$) component $\Sigma_0$ in
term of the excitation $H=({\mathcal H},{\mathcal D})$.  Substituting
(\ref{lr4}) in (\ref{em2aa}) we obtain
\begin{eqnarray}\label{dr1}
  \Sigma_0&=& -\frac c\lambda \,(-1)^{s_0+s_1}h_{\tt T}h_{\tt
    S}\,{\underline *}\,{\mathcal D}\wedge {\mathcal H}\wedge d\sigma
  \nonumber \\&&+ \frac c{2\lambda}\,(-1)^{s_0+s_1+1}{\underline
    *}\,{\mathcal D}\wedge{\mathcal D} + \frac
  1{2\lambda c}\,(-1)^{s_1}{\underline *}\,{\mathcal H}\wedge{\mathcal
    H} \,.
\end{eqnarray}
One should compare this expression with the electric current
(\ref{cur*}). We recognize in the first term the energy flux density
(or Poynting) 2-form
\begin{equation}\label{dr2}
  s:=\frac c\lambda \,(-1)^{s_0+s_1}h_{\tt T}h_{\tt S}\,{\underline
    *}\,{\mathcal D}\wedge {\mathcal H}\,.
\end{equation}
The remaining terms in (\ref{dr1}) represent the energy
density 3-forms of the electric and magnetic fields, respectively:
\begin{equation}\label{dr3}
  u_{\tt el}:=\frac c{2\lambda}\,(-1)^{s_0+s_1+1}\,{\underline *}\,{\mathcal
    D}\wedge{\mathcal D}\,, \qquad u_{\tt mg}:=\frac
  1{2\lambda c}\,(-1)^{s_1}\,{\underline *}\,{\mathcal H}\wedge{\mathcal H} \,.
\end{equation}
Thus (\ref{dr1}) can be rewritten as
\begin{equation}\label{dr4}
  \Sigma_0=-s\wedge d\sigma + u_{\tt el}+ u_{\tt mg}\,.
\end{equation}
Consequently, the relation $d\Sigma_0=0$ yields the standard form of
the continuity equation for the electromagnetic energy
\begin{equation}\label{dr5}
  {\underline d}\, s+ \frac {\partial}{\partial \sigma} \left(u_{\tt
      el}+ u_{\tt mg}\right)=0\,.
\end{equation}

Let us calculate the 3-form ${\underline *}{\mathcal H}\wedge
{\mathcal H}$ in local coordinates.  We decompose the 1-form
${\mathcal H}$ according to
\begin{equation}\label{dr6}
{\mathcal H}={\mathcal H}_1dx+{\mathcal H}_2dy+{\mathcal H}_3dz\,.
\end{equation}
Thus, 
 \begin{equation}\label{dr7}
   {\underline *}{\mathcal H}={\mathcal H}_1(-1)^{s_1}dy\wedge dz-{\mathcal
     H}_2dx\wedge dz+{\mathcal H}_3dx\wedge dy\,,
\end{equation}
and, consequently, 
 \begin{equation}\label{dr8}
   {\underline *}{\mathcal H}\wedge {\mathcal
     H}=\bigg[(-1)^{s_1}({\mathcal H}_1)^2+({\mathcal
     H}_2)^2+({\mathcal H}_3)^2\bigg] \,{}^{(3)}{\rm vol}\,.
\end{equation}
Analogously, we decompose the 2-form ${\mathcal D}$ according to
 \begin{equation}\label{dr8x}
   {\mathcal D}={\mathcal D}^1dy\wedge dz+{\mathcal D}^2dz\wedge
   dx+{\mathcal D}^3dx\wedge dy\,
\end{equation}
and obtain
 \begin{equation}\label{dr8xx}
   {\underline *}{\mathcal D}\wedge {\mathcal
     D}=(-1)^{s_1}\bigg[(-1)^{s_1}({\mathcal D}^1)^2+({\mathcal
     D}^2)^2+({\mathcal D}^3)^2\bigg] \,{}^{(3)}{\rm vol}\,.
\end{equation}
We substitute (\ref{dr8}) and (\ref{dr8xx}) in (\ref{dr3}) and find
\begin{eqnarray}\label{elenergy}
u_{\rm el}&=&\frac{(-1)^{s_0+1}\,c}{2\lambda}\, \left[(-1)^{s_1}
({\mathcal D}^1)^2+({\mathcal D}^2)^2+ ({\mathcal D}^3)^2 \right]\,^{(3)}{\rm
vol}\,,\\&&\nonumber\\
u_{\rm mg}&=&\hspace{9.5pt}\frac{(-1)^{s_1}}{2\lambda c}\, \left[(-1)^{s_1}
({\mathcal H}_1)^2+({\mathcal H}_2)^2+ ({\mathcal H}_3)^2 \right]\,^{(3)}{\rm
vol}\,.\label{mgenergy}
\end{eqnarray}
Incidentally, in SI notation $\varepsilon_0=\lambda/c$ and $\mu_0=1/
(\lambda c)$. Hence (\ref{elenergy}) and (\ref{mgenergy}) have,
indeed, the correct dimensions of energies.

For Euclidean signature with $s_0=s_1=0$, we have the electric energy
density
\begin{equation}\label{dr8xy}
  u_{\tt el}=\mathbf{-}\frac c{2\lambda}\,\bigg[({\mathcal D}^1)^2+({\mathcal
    D}^2)^2+({\mathcal D}^3)^2\bigg] \,{}^{(3)}{\rm vol}
\end{equation}
and the magnetic energy density
\begin{equation}\label{dr8xz}
  u_{\tt mg}=\hspace{11pt}\frac 1{2\lambda c}\,\bigg[({\mathcal H}_1)^2+({\mathcal
    H}_2)^2+({\mathcal H}_3)^2\bigg] \,{}^{(3)}{\rm vol} \,.
\end{equation}
Hence in the Euclidean case, the electric and magnetic energy
densities are of opposite sign. This agrees with the result of Brill
\cite{Brill}.

For Minkowskian signature with $s_0=1, s_1=0$, the electric and
magnetic energy densities are
\begin{equation}\label{dr8xxy}
  u_{\tt el}=\frac c{2\lambda}\,\bigg[({\mathcal D}^1)^2+({\mathcal
    D}^2)^2+({\mathcal D}^3)^2\bigg] \,{}^{(3)}{\rm vol}\,
\end{equation}
and
\begin{equation}\label{dr8xxz}
  u_{\tt mg}=\frac 1{2\lambda c}\,\bigg[({\mathcal H}_1)^2+({\mathcal
    H}_2)^2+({\mathcal H}_3)^2\bigg] \,{}^{(3)}{\rm vol} \,,
\end{equation}
respectively. Thus, in the Minkowskian case, the electric and magnetic
energy densities are positive provided $\lambda>0$.

For different signatures, the Poynting 2-form $s$ has different signs
and even depends on the value of the factor $h_{\tt T}h_{\tt S}$.
This does not prevent the conservation law for the total
electromagnetic energy to be formally valid for all signatures.
Indeed, it is a consequence of the four dimensional conservation law
that holds independently of the value of the factors $h_{\tt T}$ and
$h_{\tt S}$. Nevertheless, as we will see below, for Minkowskian
signature, $s$ will finally have its correct form
$s=(c/\lambda)\,\underline{*}{\mathcal D}\wedge {\mathcal H}=E\wedge
{\mathcal H}$.

\section{Field equations --- derivation of factors}

The first pair of Maxwell equations (\ref{inhomMax}) is already given
in term of the pair $({\mathcal H},{\mathcal D})$:
 \begin{equation}\label{dr8n}
   {h_{\tt T}\,\underline{d}\,{\mathcal H}+{h_{\tt
         S}}\,\frac{\partial}{\partial \sigma}{\mathcal D} = i_{\tt
       T}j\,,\qquad\quad h_{\tt S}\,\underline{d}\,\mathcal D}=-i_{\tt
     T}\rho\,.
\end{equation}
We substitute (\ref{lr4}) in the second pair of Maxwell equations
(\ref{homMax}) and obtain
\begin{equation}\label{dr9}
  (-1)^{s_0}{h_{\tt S}}\,c\,\underline{d}\, ({\underline *}\,{\mathcal D})+
   h_{\tt T}\,\frac{1}{c}\,\frac{\partial}{\partial
    \sigma}({\underline *}\,{\mathcal H})=0\,,\qquad
  \underline{d}\,({\underline *}\,{\mathcal H})=0\,.
\end{equation}
The system of these four equations determines completely ${\mathcal
  H}$ and ${\mathcal D}$ for prescribed sources $j$ and $\rho$.
Consequently, the fact that the factors $f_{\tt T}$ and $f_{\tt S}$
are no longer involved in the field equations (\ref{dr8n}) and
(\ref{dr9}) means that these constants have to be treated as
conventional.  It is natural to accept the usual convention, i.e., to
require $E$ to be in the same direction as ${\underline *}\,{\mathcal
  D}$ and similarly for $B$ and ${\underline *}\,{\mathcal H}$.  Thus,
from (\ref{lr4}), we have
\begin{equation}\label{dr10}
  f_{\tt T}=(-1)^{s_0+s_1+1}h_{\tt S}\,, \qquad f_{\tt
    S}=(-1)^{s_1+1}h_{\tt T}\,, \qquad{\rm for} \qquad \lambda>0\,.
\end{equation}
and
\begin{equation}\label{dr10x}
  f_{\tt T}=(-1)^{s_0+s_1}h_{\tt S}\,, \qquad \hspace{11pt} f_{\tt
    S}=(-1)^{s_1}h_{\tt T}\,, \hspace{11pt}\qquad{\rm for} \qquad \lambda<0\,.
\end{equation}

Observe another property of the system (\ref{dr8n}), (\ref{dr9}).  The
constant $i_{\tt T}$ appears only as a factor in front of the charge
$\rho$ and the current $j$.  We take into account that the 2-form $j$
always represents the current density of some charge.  Thus, even if
we define the 3-form $\rho$ to be positive, the charge density $i_{\tt
  T}\rho$ can be of either sign.  Thus we find: {\it Electric charges
  can be of two types: positive and negative charges.} This result is
valid for all signatures.  Incidentally, in an analogous treatment of
gravity, one finds only one type of ``charge'', namely mass-energy
with a positive sign; in contrast, negative mass-energy doesn't exist.

Now we can absorb the factor $i_{\tt T}$ into $\rho$ and $j$ or, in
other words, put $i_{\tt T}=1$. Consequently $\rho$ can carry two
opposite signs and $j$ two opposite directions. We rewrite the system
(\ref{dr8n}), (\ref{dr9}) as
\begin{equation}\label{dr11}
  \underline{d}\,{\mathcal H}+h_{\tt T}h_{\tt S}\,\frac{\partial}{\partial
    \sigma}{\mathcal D} = h_{\tt T}j\,,\qquad\quad \underline{d}\,{\mathcal
    D}=-h_{\tt S}\,\rho\,,
\end{equation}
and
\begin{equation}\label{dr12}
 c\, \underline{d}\, ({\underline *}\,{\mathcal D})+ (-1)^{s_0
     }h_{\tt T}{h_{\tt S}}\,\frac{1}{c}\,\frac{\partial}{\partial
    \sigma}({\underline *}\,{\mathcal H})=0\,,\qquad
  \underline{d}\,({\underline *}\,{\mathcal H})=0\,.
\end{equation}
The relative signs on the left-hand-sides of (\ref{dr11})$_1$ and
(\ref{dr12})$_1$, respectively, depend only on $s_0$ and not on
$h_{\tt T}{h_{\tt S}}$.  For $s_0=1$, the relative signs in
(\ref{dr11})$_1$ and (\ref{dr12})$_1$ are opposite, in
accordance with the Lenz rule.

For $s_0=0$, we have the same pair of signs --- the anti-Lenz rule.
The signature, however, is defined by two factors $s_0$ and $s_1$; for
instance, $s_0=1,\, s_1=0$ corresponds to Minkowskian signature as
well as $s_0=0,\, s_1=1$.  Thus so far we cannot establish the proper
correspondence between the signature and the sign of the induction. It
will be done in the next section by applying the expression for the
Lorentz force.
 
At this stage, we are ready to fix the factors $h_{\tt T}$ and $h_{\tt
  S}$.  The electromagnetic excitation $H$ (that is, its projections
${\mathcal H}$ and ${\mathcal D}$) can only be measured with the help of the
current $J$.  Thus, in (\ref{dr11})$_2$, the factor $h_{\tt S}$ fixes
the direction that we ascribe to the field ${\mathcal D}$ if it emanates,
say, from a positive charge. Conventionally, $\mathcal D$ is defined such
that it is directed from positive to negative charges.  Thus $h_{\tt
  S}=-1$ and (\ref{dr11}) reads
\begin{equation}\label{dr13}
  \underline{d}\,{\mathcal H}=h_{\tt T}\,\left(j+\dot{\mathcal D}\right)
  \,,\qquad \underline{d}\,{\mathcal D}=\rho\,.
\end{equation}
Eq.(\ref{dr13}) represents the Amp\`ere-Maxwell law.  Note that the
contributions of $j$ and $\dot{\mathcal D}$ already emerge with the
correct relative sign. In analogy to $\mathcal D$, the magnetic excitation
$\mathcal H$ is defined such that it has $j+\dot{\mathcal D}$ as source, that
is, $h_{\tt T}=1$. Accordingly, the  Maxwell
equations (\ref{dr11}), (\ref{dr12}) finally read 
\begin{equation}\label{dr14}
  \underline{d}\,{\mathcal H}-\,\frac{\partial}{\partial \sigma}{\mathcal
    D}=j\,,\qquad\quad \underline{d}\,{\mathcal D}=\rho\,,
\end{equation}
and
\begin{equation}\label{dr15}
 c\, \underline{d}\, ({\underline *}\,{\mathcal D})+
  (-1)^{s_0}\,\frac{1}{c}\,\frac{\partial}{\partial \sigma}({\underline *}\,{\mathcal
    H})=0\,,\qquad \underline{d}\,({\underline *}\,{\mathcal H})=0\,.
\end{equation}

We collect the sign factors the values of which have already been set:
\begin{equation}\label{dr15*}
  i_{\tt T}=1\,,\quad i_{\tt S}=-1\,,\quad h_{\tt T}=1\,,\quad h_{\tt
    S}=-1\,.
\end{equation}
The sign factors of the field strength $F$, in our convention
(\ref{dr10}), depend on the signature {\em and\/} on the sign of
$\lambda$.  For $\lambda>0$, we listed the values for the different
signatures in Table~2.

\begin{table}[htbp]
  Table 2. The sign factors $f_{\tt T},\,f_{\tt S}$ of the field
  strength $F$ for different signatures in the case of
  $\lambda>0$\medskip
\begin{center}
{\begin{tabular}{||c||c|c||}
\hline\hline
Signature&$f_{\tt T}$&$f_{\tt S}$\\ \hline\hline
Minkowski (aka Lorentz) & -1 & -1  \\ \hline
Euclid & 1 & -1 \\ \hline
$(-,-,+,+)$ &  1 & 1 \\ \hline\hline
\end{tabular}}
\end{center}
\end{table}

\section{Lorentz force --- relation between different signatures}

The components of the Lorentz force density (\ref{lor5}) and
(\ref{lor6}), for $i_{\tt T}=1$, see (\ref{dr15*}), can be rewritten
according to
\begin{equation}\label{lor5@}
  f_0=-f_{\tt T}\,E\wedge j\wedge d\sigma\,,\qquad f_\mu=\bigg[f_{\tt
    T}(e_\mu\rfloor E)\rho+ f_{\tt S}(e_\mu\rfloor B)\wedge
  j\bigg]\wedge d\sigma\,.
\end{equation}
We substitute (\ref{lr4}) in (\ref{lor5@}) and use the already fixed
values (\ref{dr15*}) of the factors $h_{\tt T}, h_{\tt S}$, and
$i_{\tt T}$. Then,
\begin{equation}\label{intr1}
  f_0=-\frac c\lambda\, (-1)^{s_0+s_1}\,{\underline *}\,{\mathcal D}\wedge
  j\wedge d\sigma\,
\end{equation}
and 
\begin{equation}\label{intr2}
  f_\mu=\frac 1\lambda\bigg[(-1)^{s_0+s_1}\,c\, (e_\mu\rfloor
  {\underline *}\,{\mathcal D}) \rho+
  (-1)^{s_1+1}\,\frac{1}{c}\,(e_\mu\rfloor {\underline *}\,{\mathcal
    H})\wedge j\bigg]\wedge d\sigma\,.
\end{equation}
The latter expression represents the ordinary electromagnetic 3-force.
Using the identity $e_\mu\rfloor *w=*(w\wedge dx_\mu)$, we can rewrite
it as
\begin{equation}\label{intr3}
  f_\mu=\frac 1\lambda\bigg[(-1)^{s_0+s_1}\,c\, {\underline
    *}\,\rho\,{\mathcal D}+ (-1)^{s_1}\,\frac{1}{c}\, {\underline
    *}\,j\wedge{\mathcal H}\bigg]\wedge dx_\mu\wedge d\sigma\,.
\end{equation}

Let us examine this expression for the different signatures. For
static fields ${\mathcal D}$ and ${\mathcal H}$, the signature is completely
eliminated from the field equations (\ref{dr14}), (\ref{dr15}).  Thus,
for given sources $\rho$ and $j$, we have the same static fields for
all signatures.

$\bullet$ 
For $s_0=1,\, s_1=0$, we have the ordinary Maxwell-Lorentz
electrodynamics in Riemannian spacetime with Minkowskian signature.
The first term on the right-hand-side of (\ref{intr3}) represents the
electric force between two static charges. In particular, it yields
attraction between opposite charges and repulsion between charges of
the same sign: this is Dufay's law.  The second term describes the
magnetic force.  In particular, it is responsible for the pulling of a
ferromagnetic core into a solenoid independently on the direction of
the current, in accordance with Lenz's rule.

$\bullet$ 
For $s_0=0,\, s_1=0$, we deal with Euclidean electrodynamics.  In
contrast to the case of Minkowskian signature, the energy density of
the electromagnetic field does not have a definite sign.  Hence also
the sign of $\lambda$ is not defined.

In the case $\lambda>0$, the Euclidean electric term in (\ref{intr3})
is opposite to the corresponding term of ordinary (Minkowskian)
electrodynamics.  Consequently, we obtain an anti-Dufay law: opposite
charges repel whereas charges of the same sign attract each other. As
for the Euclidean magnetic force, it comes with the same sign as in
ordinary electrodynamics, in accordance with the Lenz rule.


In the case $\lambda<0$, the situation is opposite: 
the Dufay law for charges and the anti-Lenz rule for currents.

These results are in correspondence with the signs of the Euclidean
electric and magnetic energy densities (\ref{dr8xy},\ref{dr8xz}).
They are partly presented in Brill's analysis \cite{Brill}.

$\bullet$ For $s_0=1,\, s_1=1$, i.e., for the signature $(-,-,+,+)$,
the two factors in (\ref{intr3}) have opposite signs relative to the
Minkowskian case.  Thus, for $\lambda>0$, we have anti-Dufay and
anti-Lenz laws.  For $\lambda<0$, the laws are the same as in ordinary
electrodynamics.  A surprising fact is that the forces have definite
signs, although the signs of the energy densities are undefined.

\section{Main results and discussion}

Let us recall the main points of our analysis:
\begin{itemize}
\item[(i)] We formulate, in a metric-free form, the field equations,
  the Lorentz force, and the conservation law for energy-momentum.
\item[(ii)] The spacetime relation is assumed to be local, linear, and
  symmetric.
\item[(iii)] The field equations and the Lorentz force are shown to be
  compatible with a certain energy-momentum current $\Sigma_\a$.
\item[(iv)] In order to provide a physical interpretation of the
  4-dimensional quantities, we construct their $(1+3)$-decompositions
  with a number of free sign factors.
\item[(v)] We observe a specific symmetry of the energy-momentum
  current $\Sigma_\a$, namely electric/magnetic reciprocity. In
  $(1+3)$-decomposition it translates into an exchange between
  electric and magnetic fields.
\item[(vi)] If electric/magnetic reciprocity is applied to the
  spacetime relation, see (ii), it yields a metric of Lorentzian
  type.
\item[(vii)] For all possible signatures of the 4-dimensional metric,
  we obtain the expressions for the electric and magnetic energy
  densities.  The metric of a Lorentzian type turns out to be related
  to a positive electromagnetic energy density.  This result does not
  depend on the values of the sign factors.
\item[(viii)] We analyze the $(1+3)$-decompositions of the field
  equations. We derive which sign factors are conventional and which
  do depend on the signature.  We find that the electric charge has
  two possible signs for all signatures.
\item[(ix)] For all signatures, we derive the features of the
  interactions between charges (Dufay's law) and between currents
  (Lenz's rule).
\end{itemize}

All in all, our discussion has shown that the {\em Lorentzian\/}
signature of the metric of spacetime originates in properties of the
electromagnetic spacetime relation $H\!=\!H(F)$. If the spacetime
relation is local, linear, and symmetric, the requirement of
electric/magnetic reciprocity induces the lightcone (with Lorentzian
signature) and a positive definite electromagnetic energy together
with Dufay's law and Lenz's rule.

\section*{Acknowledgments}
We are grateful to the Erwin Schr\"odinger International Institute for
Mathematical Physics (ESI) for its hospitality. Furthermore we would
like to thank Wolfgang Kummer of the Technical University of Vienna as
well as his co-organizers D.\ Grumiller, H.\ Nicolai, and D.\ 
Vassilevich for the invitations to participate at the Workshop on 2D
gravity at the ESI in Vienna. Yuri Obukhov read our paper very
carefully and came up with many questions and remarks. We both thank
him very much.


\end{document}